\documentclass{svproc}
\usepackage{url}

\usepackage{graphicx}
\usepackage{amsmath,amssymb}

\usepackage{caption}
\usepackage{booktabs}
\usepackage{xcolor}

\usepackage{array}

\newcommand{\envelope}{\textsuperscript{$\star$}}

\newcommand{\qmax}{q_{\max}}

\usepackage[hidelinks]{hyperref}
\begin{document}
\mainmatter

\title{CORTET: Robust generation of simulation-ready {tetrahedral meshes of the fetal cerebral cortex}}

\titlerunning{CORTET: Automated Cortical Meshing}

\author{Davood Shahsavari\inst{1,3}\thanks{Corresponding authors: \email{davood.shahsavari@kcl.ac.uk}, \email{emma.robinson@kcl.ac.uk}} \and Irina Grigorescu\inst{1} \and Katherine R. Long\inst{3,4} \and
Jiaxin Xiao\inst{2} \and Kaili Liang \inst{1,2,5} \and Nashira Baena\inst{1} \and Aakash Saboo\inst{1} \and  Saga N.B. Masui\inst{1} \and Yourong Guo\inst{1}  \and Vanessa Kyriakopoulou\inst{2} \and
Alena Uus\inst{2} \and
Martin J. Bishop\inst{1}  \and Emma C. Robinson\inst{1,2}\envelope}

\authorrunning{D. Shahsavari et al.}

\tocauthor{Davood Shahsavari, Irina Grigorescu, Katherine R. Long, Jiaxin Xiao, Kaili Liang, Nashira Baena, Aakash Saboo, Saga N.B. Masui,  Yourong Guo, Vanessa Kyriakopoulou, Alena Uus,
Martin J. Bishop, Emma C. Robinson}

\institute{%
Research Department of Biomedical Computing,
School of Biomedical Engineering \& Imaging Sciences, King's College
London, UK\\ 
\and Research Department of Early Life Imaging, School of Biomedical Engineering
\& Imaging Sciences, King's College London, UK
\and Centre for Developmental Neurobiology, Institute of Psychiatry,
Psychology \& Neuroscience, King's College London, UK
\and MRC Centre for Neurodevelopmental Disorders, King's College
London, UK\\
\and Department of Child and Adolescent Psychiatry, Institute of Psychiatry, Psychology \& Neuroscience, King’s College London, London, SE5 8AF, UK}

\maketitle

\begin{abstract}
Every human brain folds differently, and such natural variation confounds the search for imaging biomarkers of neurodevelopmental disorders. Physics-based simulation can help determine the causal mechanisms that underpin this variability.  Yet every simulation must be initiated from a {volumetric} mesh of the brain's interior{, tetrahedral or hexahedral}, and it is the worst element in that mesh, not the average, that decides whether the simulation runs at all. Building that mesh  from fetal MRI currently requires labour-intensive manual intervention. We therefore present CORTET (CORtical TETrahedral meshing): a fully automated pipeline that converts a triangulated cortical surface into a solver-ready tetrahedral mesh whose worst-element quality meets a strict {quality} target with no manual repair. {By benchmarking against a general-purpose tetrahedral mesher on the same input surfaces, we isolate the pipeline's contribution from that of the input geometry, and we validate quality across a cohort of nearly 200 fetal subjects spanning the folding period.} A mesh taken straight from the pipeline sustains a numerically stable morphoelastic folding simulation of a real fetal subject.
\end{abstract}

\section{Introduction}

Over the third trimester the fetal cortex folds, driven by mechanical stress as the cortical grey matter expands faster than the tissue beneath it. 
When this process goes wrong, the consequences are severe and well characterised. These include malformations of cortical development such as lissencephaly, the absence of folds, and polymicrogyria, an excess of small ones~\cite{bayly2014review}. Altered cortical folding has also been reported in epilepsy, autism, schizophrenia~\cite{bayly2014review,kyriakopoulou2014}, preterm birth~\cite{dasilva2025}, and congenital heart disease~\cite{ortinau2019}.
{Explaining the physics of cortical folding therefore offers a route towards understanding these conditions. However, cortical folding involves a large-deformation buckling instability, making its numerical simulation susceptible to failure. Common failure modes include self-intersection of the deforming mesh and severe element distortion leading to element inversion \cite{bayly2013,budday2014,tallinen2014}. These are artefacts arising from the discretisation rather than features of the tissue’s physical behaviour. The quality of the initial mesh is therefore a major determinant of whether the simulation can be completed successfully.}

Finite-element models have established the mechanical basis of folding: a soft, growing cortex constrained by a stiffer core buckles into folds ~\cite{bayly2013,tallinen2014,tallinen2016}. The whole-brain model of Tallinen et al.~\cite{tallinen2016}, initialised from a 22-week fetal MRI scan, produced folds with realistic shape and wavelength; yet their location and orientation do not match those of the human brain.
The picture has since grown richer: growth couples back to the tissue it deforms, with folding stresses reorganising subcortical white-matter fibres, growth rates vary spatially across the cortex~\cite{garcia2018dynamic}, and cortical thickness and growth set the fold wavelength~\cite{budday2014,wang2021}. Each refinement makes the model more faithful  and more patient-specific, and so raises the same practical demand: to test these mechanisms one must simulate many individual brains from their own imaging. This is now becoming feasible as open longitudinal fetal MRI datasets appear, yet simulations still run on idealised geometries~\cite{darayi2022} or a single template brain,
{and a patient-specific fetal folding pipeline required manual re-meshing in 10 of 29  cases ~\cite{alenya2022}. This process involved iteratively removing or smoothing problematic elements and correcting topological defects, such as handles and holes, until the solver converged.}

A major obstacle to current progress is the question of how to robustly generate simulation-ready tetrahedral meshes from cortical surfaces. {The difficulty is anatomical rather than incidental. A fetal cortical surface is a closed surface of genus zero, but one that folds into gyri and sulci as gestation proceeds, and the sulci are both deep and narrow: opposing banks approach one another far more closely than the scale of the structure they sit in, and the local curvature at a sulcal fundus is correspondingly high. A tetrahedraliser must therefore resolve the separation between two facing banks while filling a volume orders of magnitude larger, and it is inside those constrictions that flat, near-degenerate elements appear. The constraint tightens with gestational age as folding deepens, which is why it is the latest ages that bound refinement (Sec.~\ref{sec:discussion}).}
General-purpose tetrahedralisers, TetGen~\cite{si2015tetgen}, Netgen~\cite{schoberl1997netgen}, CGAL~\cite{cgal_3d_mesh}, fail to generate meshes that run mechanical simulations to completion; largely because {the meshes they produce contain many tetrahedra of near-zero volume}.
Meshing directly from labelled image volumes~\cite{lederman2011,fang2009} inherits voxel staircase artifacts, requiring surfaces to be smoothed, erasing the sub-millimetre detail that cortical folding depends on.
What is missing is an automated pipeline that takes an individual cortical surface and pushes it through tetrahedralisation and regularises the output in a way that guarantees that the \emph{worst} element, not merely its average, is good enough to fold, without a person in the loop.

\paragraph{Contributions.} In this paper, we present (1) CORTET (CORtical TETrahedral meshing), a fully automated pipeline that turns an individual cortical surface into a solver-ready tetrahedral mesh with worst-element quality $\qmax<0.6$ {(each element carries a score $q_e\in[0,1]$, where $0$ is a regular tetrahedron and $1$ a degenerate one, and $\qmax=\max(q_e)$ is the worst element in the mesh; Sec.~\ref{sec:eval})} and no manual repair; and (2) perform validation across a cohort of nearly 200 fetal subjects spanning the folding period, inclusive of an ablation study that isolates the contribution of each step of the pipeline; in order to (3) sustain a numerically stable folding simulation of a real fetal subject.

\section{Modelling framework}

{This mechanical formulation determines the requirements placed on the simulation mesh.} Cortical folding is modelled as morphoelastic growth in which the deformation gradient splits multiplicatively, $\mathbf{F}=\mathbf{F}_e\mathbf{F}_g$, into a stress-free tangential growth component $\mathbf{F}_g$ and an elastic component $\mathbf{F}_e$ that generates stress through a neo-Hookean response~\cite{rodriguez1994}. Folds set in when the cortex--core stiffness contrast and accumulating growth drive a buckling instability, with fold wavelength scaling as $\lambda\sim H\,(\mu_c/\mu_s)^{1/3}$ for cortical thickness $H$ and cortex-to-core shear-modulus ratio $\mu_c/\mu_s$~\cite{bayly2013,tallinen2014,shahsavari2025surface}.

Such models are solved {here} by explicit time integration, which makes mesh quality decisive in two ways. First, each element's deformation gradient is {computed} by inverting its undeformed edge matrix; when a tetrahedron degenerates, its vertices become nearly coplanar, this matrix becomes near-singular, and the resulting Jacobian $J=\det\mathbf{F}$ can turn non-positive, inverting the element and halting the solver.
A single degenerate element is therefore enough to stop a run, so the \emph{worst} element, not the average, decides whether a mesh is usable. {This first requirement is not specific to explicit integration: $J=\det\mathbf{F}$ must be evaluated for any hyperelastic model, and a degenerate element ill-conditions
the tangent stiffness of an implicit solver just as readily, where the consequence is a failure to converge rather than an immediate halt. Only the second consideration below is specific to explicit integration.} Second, the stable time step is bounded by $\Delta t_{\mathrm{crit}}\le \ell_{\min}/c$, where $\ell_{\min}$ is the shortest edge and $c$ the elastic wave speed, so one short edge throttles the whole simulation. Worst-element quality and edge uniformity are thus the two properties a mesh generator must control.

{The discretisation follows the same model reported by \cite{tallinen2014,tallinen2016}: linear four-node, constant-strain tetrahedra advanced by an explicit solver. Brain tissue is usually modelled as nearly incompressible~\cite{millerchinzei1997}, with a Poisson ratio of 0.45--0.49 recommended for slow processes such as development~\cite{budday2020review}, and linear tetrahedra lock volumetrically in that limit~\cite{bonet1998simple}. The solver therefore takes the pressure from a nodally averaged Jacobian, the average nodal pressure element~\cite{bonet1998simple}, which with a single bulk modulus coincides with the non-locking element of~\cite{joldes2009nonlocking}. The run of Sec.~\ref{sec:res-sim} keeps the bulk modulus of the published model, $K=5\mu$ (Poisson ratio $\approx0.4$)~\cite{tallinen2014}, so that it can be compared with that model. We use tetrahedra rather than hexahedra because Delaunay refinement conforms automatically to a deeply folded cortical surface, while a conforming hexahedral mesh of the same geometry is far harder to automate~\cite{wittek2016meshes}.}

\section{Methods}

\subsection{Pipeline overview}

\begin{figure}[htbp]
\centering
\includegraphics[width=0.83\textwidth]{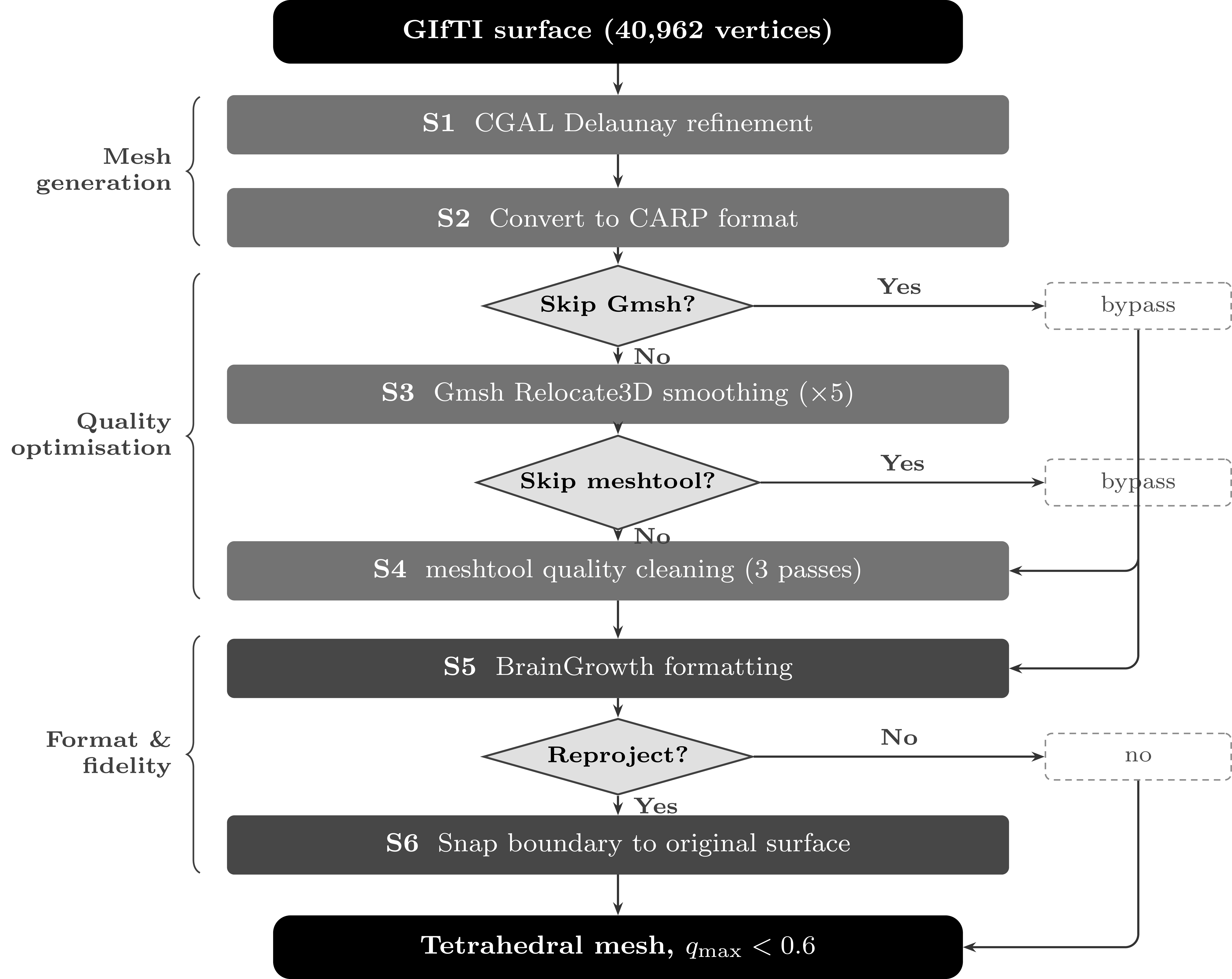}
\caption{The six-stage pipeline from a GIfTI cortical surface to a
finite-element-ready tetrahedral mesh. Diamonds (stages 3, 4, 6) are optional bypass
gates for matching the output to a target solver.}
\label{fig:pipeline}
\end{figure}

CORTET converts a closed, orientable triangulated cortical surface into an optimised tetrahedral volume mesh through six sequential stages, grouped by function: stages S1--S2 generate the volume mesh, S3--S4 optimise its quality, and S5--S6 format the output and control surface fidelity (Fig.~\ref{fig:pipeline}). Stages S3, S4 and S6 are optional and can be bypassed so the output matches a given target solver. 
The pipeline is implemented in Python and chains four established tools, each in one role: CGAL~\cite{cgal_3d_mesh}, a computational-geometry library, builds the tetrahedralisation via its \texttt{pygalmesh} bindings~\cite{pygalmesh}; Gmsh~\cite{gmsh} applies global vertex smoothing; and \texttt{meshtool}~\cite{meshtool} performs worst-element cleaning; with NiBabel~\cite{nibabel}, meshio~\cite{meshio} and Trimesh~\cite{trimesh} handling input and format conversion. We assume surfaces in GIfTI format, which lets us resample meshes and interpolate cortical features with the Connectome Workbench~\cite{marcus2011workbench}.

\subsection{Volume generation and quality optimisation (Stages 1--4)}

CGAL's Delaunay refinement fills the surface with a tetrahedral volume, {and the process starts with using it}. Because it assumes that the interior of the mesh is identified from the surface's face normals, we first repair inconsistently oriented triangles and export a temporary STL. During this process, a single parameter, the cell size $h$ {(the characteristic element dimension: the upper bound CGAL places on each tetrahedron's circumradius and on the surface Delaunay ball radius, with the facet-distance tolerance set to $h/2$)}, sets element density. CGAL then applies four optimisation passes: the first two--Optimal Delaunay Triangulation and Lloyd~\cite{lloyd1982}--work to relocate interior vertices so as to regularise element size and shape; the third perturbs the worst-shaped elements; then a final pass removes sliver elements (whose four vertices are almost coplanar) by reweighting vertices in place, which is why it runs last {(S1)}. The refinement replaces every input vertex with a new node set, so per-vertex surface fields (thickness, curvature, growth rate) must be transferred to the volume by barycentric interpolation {(S2)}. Two tools then optimise quality in a complementary way: Gmsh Relocate3D applies global vertex smoothing to reduce non-uniformity of volume elements {(S3)}, and \texttt{meshtool} {improves} the worst elements through three passes of progressively tightened quality thresholds {(S4)}{, applied at $q>0.7$, $0.5$ and $0.3$ respectively}. {Each pass shifts the vertices of elements scoring worse than the threshold; element count and connectivity are unchanged. No element is discarded and no region is re-tetrahedralised, which is how the worst elements are improved without disturbing the mesh the earlier stages have already optimised.} Acting globally then locally, and tightening in stages rather than in one aggressive pass, avoids creating new poor elements while fixing others.

\subsection{Solver-ready output (Stages 5--6)}
Stage 5 writes the mesh in the target solver's format.  For an explicit growth solver this requires reconstructing a boundary-face list from the tetrahedral connectivity, ordering the surface nodes contiguously so that surface-indexed fields resolve correctly, and matching the solver's node and sign conventions, none of which are reported by the mesh generator and each of which, if violated, allow simulations to run to completion while demonstrating corrupted mechanics, rather than raising an error {(S5)}.
{Routine verification of an explicit run, such as inspection of the energy--time history, is directed at instability and does not reveal these faults: they alter the direction in which elements grow rather than the stability of the relaxation, and in a damped quasi-static formulation the run remains well behaved either way. We therefore verify these conventions on the mesh itself, before the solver is called.}
An optional Stage 6 snaps each boundary node onto the nearest point of the input surface when exact overlay with imaging is required, re-checking for inversion afterwards {(S6)}. Because the core stages are solver-agnostic, the mesh can also be exported to Abaqus, FEniCS, VTK or Gmsh formats, so that it can be used in other solvers such as FEBio.

\subsection{Data}
\label{sec:data}

We use over $200$ white-matter (inner cortical) surfaces derived from individual fetal subjects of the developing Human Connectome Project (dHCP). The surfaces were generated from T2-weighted MRI acquired on a 3T Philips Achieva system with a 32-channel cardiac coil (TE $=250$~ms, $1.1\times1.1$~mm in-plane resolution, $2.2$~mm slice thickness)~\cite{price2019}, reconstructed from 2D stacks to a coherent 3D volume, at $0.5$~mm isotropic resolution, using slice-to-volume reconstruction.  Cortical surfaces were then produced with CoTAN~\cite{ma2023conditional,grigorescu2026sud}, which takes the reconstructed volumes as input and trains a convolutional U-Net to learn a stationary velocity field that deforms a template mesh to each subject's inner cortical boundary.

\subsection{Pre-processing}
\label{sec:preproc}

Following surface fitting, Multimodal Surface Matching (MSM)~{\cite{robinson2014msm,robinson2018,besenczi2024high}} aligns all individuals to a population-average template~\cite{bozek2018}, allowing a single template mesh topology to be resampled onto each native surface, while enforcing vertex correspondence across subjects for later integration of population priors and cortical features.
Because MSM performs spherical registration, templates may be defined from sixth-order icospheric tessellations ($40{,}962$ vertices) that are subsequently resampled from the sphere onto the cortical anatomy with \texttt{wb\_command -surface resample}~\cite{marcus2011workbench}.

Prior to meshing, each aligned surface is smoothed. {We use PyMeshLab's \texttt{apply\_coord\_laplacian\_smoothing\_surface\_preserving}, a surface-preserving Laplacian smoothing filter, which constrains vertex displacement by the resulting change in face-normal orientation~\cite{muntoni2021pymeshlab}.} The required amount of smoothing is subject-dependent: younger cortices, whose folds are still forming, tolerate more smoothing, whereas older, more folded cortices deflate rapidly and require far less. We therefore set the number of smoothing iterations per subject by gestational age {(GA)}, subject to a limit on cortical-area loss and on surface displacement; we confirm preservation by tracking curvature and area before and after.

\subsection{Evaluation framework and metrics}
\label{sec:eval}

\paragraph{Choosing a quality measure.}Tetrahedral element quality can be quantified by several established shape measures, the mean-ratio~\cite{liujoe1994}, radius-ratio, volume-to-edge and radius-edge metrics~{\cite{liujoe1994,shewchuk2002,knupp2001algebraic}}, none of which is uniquely canonical.
{We therefore require a measure that identifies the single worst element and is consistent with the criterion used by the mesh-repair pipeline, rather than one that describes average element shape. This is because our objective is to certify that a mesh will not cause an explicit solver to stall. By \emph{stalling}, we mean a run that terminates before the prescribed growth is reached, through element inversion or through collapse of the stable time step.}

We therefore adopt \texttt{meshtool}'s \texttt{tet\_qmetric\_volume}, the measure by which the Stage-4 cleaning identifies and iteratively improves the poorest elements.
It is a volume-based per-element score normalised to $[0,1]$, on which $q_{\mathrm{e}}=0$ is a regular (equilateral) tetrahedron and $q_{\mathrm{e}}\to1$ a degenerate, flat element of vanishing volume.
{ It is defined operationally by \texttt{meshtool}'s implementation. That is also the reason we adopt it, since the Stage-4 cleaning acts on the same criterion. Reporting the measure the pipeline itself optimises avoids certifying a mesh with one metric while repairing it against another (agreement with established literature measures is tested numerically by \cite{liujoe1994} in Section \ref{sec:res-resolution}).}
For every mesh, we report the full distribution of $q_{\mathrm{e}}$, its minimum, maximum, mean $\bar q$ and standard deviation, but emphasise the worst-element: $\qmax=\max(q_{\mathrm{e}})$, since a single degenerate element is sufficient to halt the solver. 
We adopt $\qmax<0.6$ as the solver-ready criterion, a conservative margin below the $q\approx0.9$ inversion regime observed in practice.

\paragraph{Experiments.} We report three experiments. A \emph{stage ablation} decomposes the optimisation {across the full cohort}, isolating each stage's contribution on identical connectivity (Sec.~\ref{sec:res-ablation}). 
These results are benchmarked against out-of-the-box meshing {with} a general-purpose tetrahedraliser, TetGen~\cite{si2015tetgen}, and against the published Tallinen 22-week fetal mesh~\cite{tallinen2016}. A \emph{cohort study} meshes {every subject} at the default resolution, testing whether worst-element quality is maintained as the cortex grows and folds (Sec.~\ref{sec:res-cohort}).
An \emph{effect-of-resolution} study varies the cell size $h$ on {three subjects} to characterise the resolution--quality trade-off (Sec.~\ref{sec:res-resolution}).

\section{Results}
\label{sec:results}

\subsection{Pipeline performance {across the cohort and} stage contributions}
\label{sec:res-ablation}

\begin{figure}[t]
  \centering
  
  \includegraphics[width=\textwidth, height=6cm]{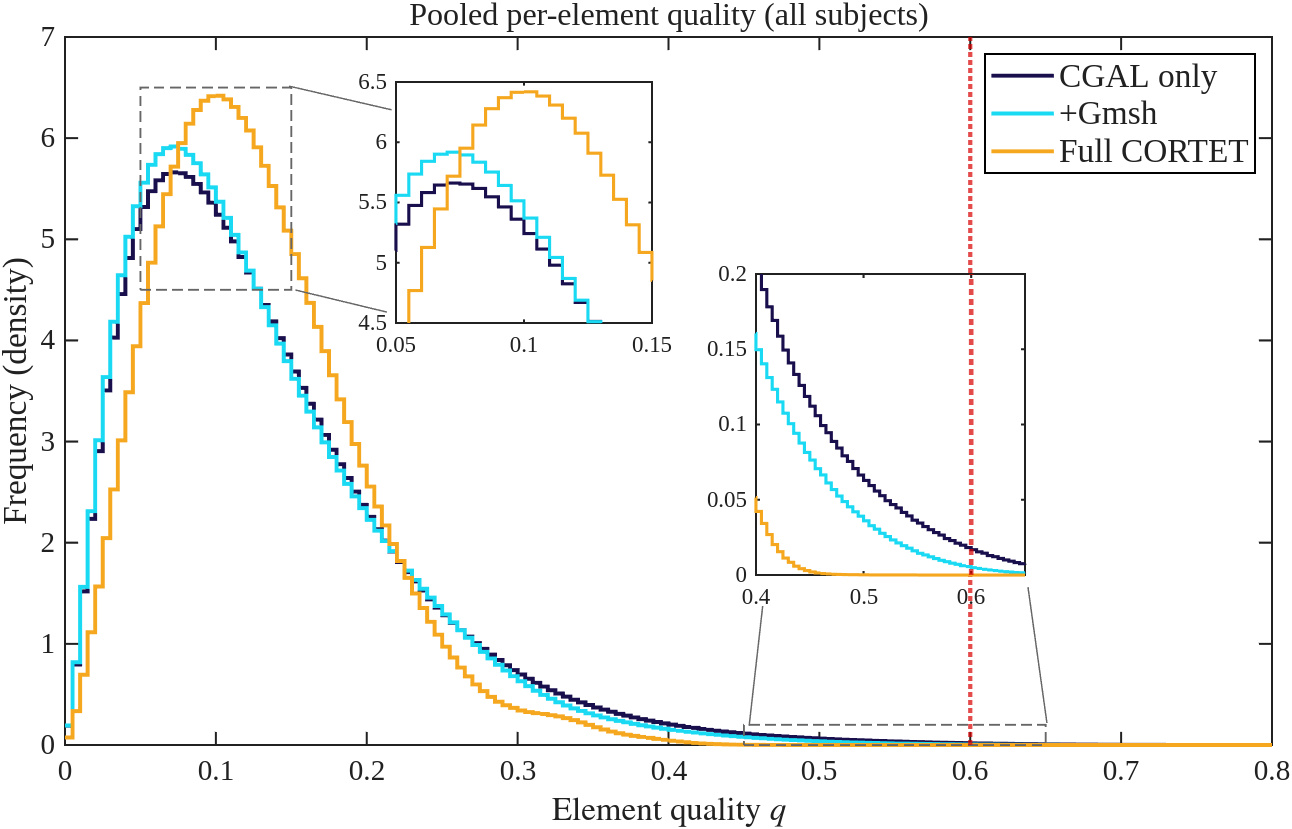}
  \caption{Per-element quality pooled across the cohort. The dotted line marks the solver-ready threshold
    $q=0.6$. {Configurations: CGAL only (S1--S2), + Gmsh (S1--S3), and full CORTET with meshtool cleaning (S1--S4)}.}
  \label{fig:pooled_hist}
\end{figure}

{
We evaluate the pipeline on $N=194$ left-hemisphere subjects spanning gestational weeks 21--38, meshing each at the default resolution ($h=0.6$~mm) and measuring every element with \texttt{tet\_qmetric\_volume} (Table~\ref{tab:ablation} and Fig.~\ref{fig:pooled_hist}).

The result is unequivocal. CGAL refinement alone attains a good \emph{mean} element quality ($\bar q = 0.135 \pm 0.001$), yet every one of the 194 subjects carries elements deep in the degenerate regime: a median of 888 elements per mesh exceed the solver-ready threshold $q>0.6$ (range 274--2{,}449), with a median worst element of $\qmax = 0.780$. 
Global Gmsh smoothing improves the distribution but does not resolve it, the median count of
threshold-exceeding elements falls to 191 and $\qmax$ to $0.732$, yet no mesh becomes compliant. Only after \texttt{meshtool} cleaning does the tail vanish entirely: across all tetrahedra in the cohort, \emph{not one element} exceeds $q=0.6$, and no subject's worst element exceeds $0.561$ (median $0.508$, range $0.462$--$0.561$). 
Every subject meets the criterion, without per-subject tuning, despite meshes differing by an order of magnitude in size and cortices ranging from smooth to deeply folded. This exposes why the mean cannot certify a mesh for simulation. Across configurations, mean quality varies by less than $0.01$ ($0.135 \rightarrow 0.129$) while the number of solver-breaking elements falls from $1.7\times10^{5}$ to zero. 
The pooled distributions (Fig.~\ref{fig:pooled_hist}) make this concrete: in bulk they are nearly indistinguishable, the full pipeline merely shifting and sharpening the mode; in the tail they diverge decisively, with CGAL and Gmsh leaving a population of elements beyond the threshold while the full pipeline terminates well before it.

The gain is attributable to the pipeline rather than to the input geometry. Run out-of-the-box on the \emph{same} surfaces, a general-purpose tetrahedraliser (TetGen) leaves a median of 15{,}955 threshold-exceeding elements per mesh (range 12{,}095--35{,}414) and reaches a fully degenerate worst element ($\qmax = 1.000$), failing on every subject. {To test whether the same worst-element guarantee could be obtained from a general-purpose tetrahedraliser given identical downstream processing, we applied Stage~3 (Gmsh) and Stage~4 (meshtool) to a TetGen mesh of the same subject. Worst-element quality remained effectively unchanged ($\qmax = 0.980 \to 0.9999$), in contrast to the $\sim$30--40\% reduction the same stages achieve on CGAL's output ($\qmax = 0.75$--$0.82 \to 0.46$--$0.56$), indicating the guarantee depends on properties of the CGAL tetrahedralisation itself rather than the cleaning stages alone.}
Moreover, for reference, the published 2016 fetal mesh~\cite{tallinen2016}, which is known to have sustained morphoelastic folding simulations, retains 15 such elements ($\qmax = 0.811$). 
Our criterion is therefore stricter than the standard met by a mesh that folds in practice, and the pipeline meets it with margin.}

\begin{table}[t]

\caption{Mesh quality across the cohort. {Stage labels refer to Fig.~\ref{fig:pipeline}.} Mean quality $\bar q$ is reported as mean $\pm$ s.d. over subjects; worst-element quality $\qmax$ and the per-subject count of elements exceeding the solver-ready threshold $q>0.6$ are reported as median [min--max]. TetGen is run out-of-the-box on the same input surfaces.
The Tallinen 22-week mesh is a single published reference. Only the full pipeline leaves no element above the threshold, in any subject. Lower $q$ is better.}
\label{tab:ablation}
\centering
\small
\setlength{\tabcolsep}{6pt}
\begin{tabular}{@{}l c c c@{}}
\toprule
Configuration & $\bar q$ & $\qmax$ & Elements $q>0.6$ \\
\midrule
TetGen  & 0.242 $\pm$ 0.005 & 0.996 [0.974--1.000] & 15{,}955 [12{,}095--35{,}414] \\
Tallinen 22W & 0.162 & 0.811 & 15 \\
\midrule
CGAL only {(S1--S2)}      & 0.135 $\pm$ 0.001 & 0.780 [0.746--0.820] & 888 [274--2{,}449] \\
\;+ Gmsh {(S1--S3)}       & 0.129 $\pm$ 0.001 & 0.732 [0.685--0.802] & 191 [55--519] \\
Full CORTET {(S1--S4)}    & 0.129 $\pm$ 0.001 & 0.508 [0.462--0.561] & \textbf{0 [0--0]} \\
\bottomrule
\end{tabular}
\end{table}

\subsection{Quality across the cohort and development}
\label{sec:res-cohort}
\begin{figure}[t]
  \centering
   
   \includegraphics[width=\textwidth, height=4.5cm]{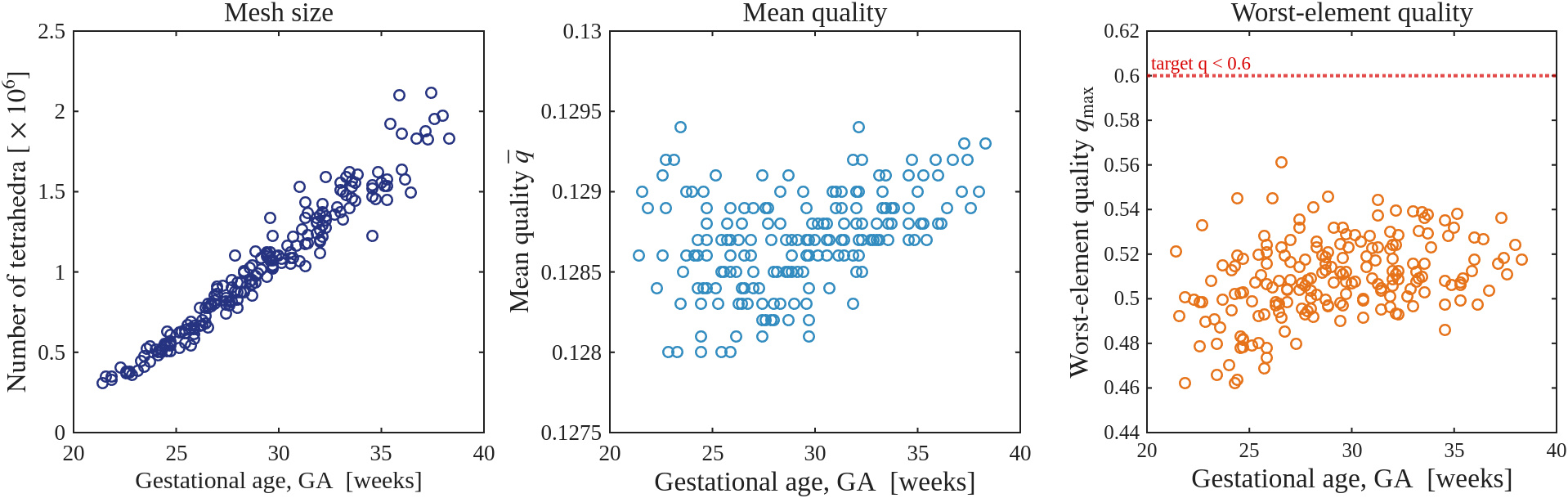}
\caption{Cohort mesh quality across gestational age ($194$ subjects, left hemisphere, $h=0.6$~mm), shown as three single-axis panels against gestational age (GA).
\textbf{Left:} tetrahedron count, which grows from $\sim\!0.3$ to $\sim\!2.1$~million as the cortex enlarges and folds. \textbf{Centre:} mean element quality $\bar q$, essentially constant at $\approx0.129$ across gestation.
\textbf{Right:} worst-element quality $\qmax$, which remains below the $\qmax<0.6$ target (dotted red) for every subject. Quality (\texttt{tet\_qmetric\_volume}, 0 = ideal) is thus maintained independently of mesh size and fold complexity.}
  \label{fig:cohort}
\end{figure}

\begin{figure}[t]
\centering
\includegraphics[width=\linewidth]{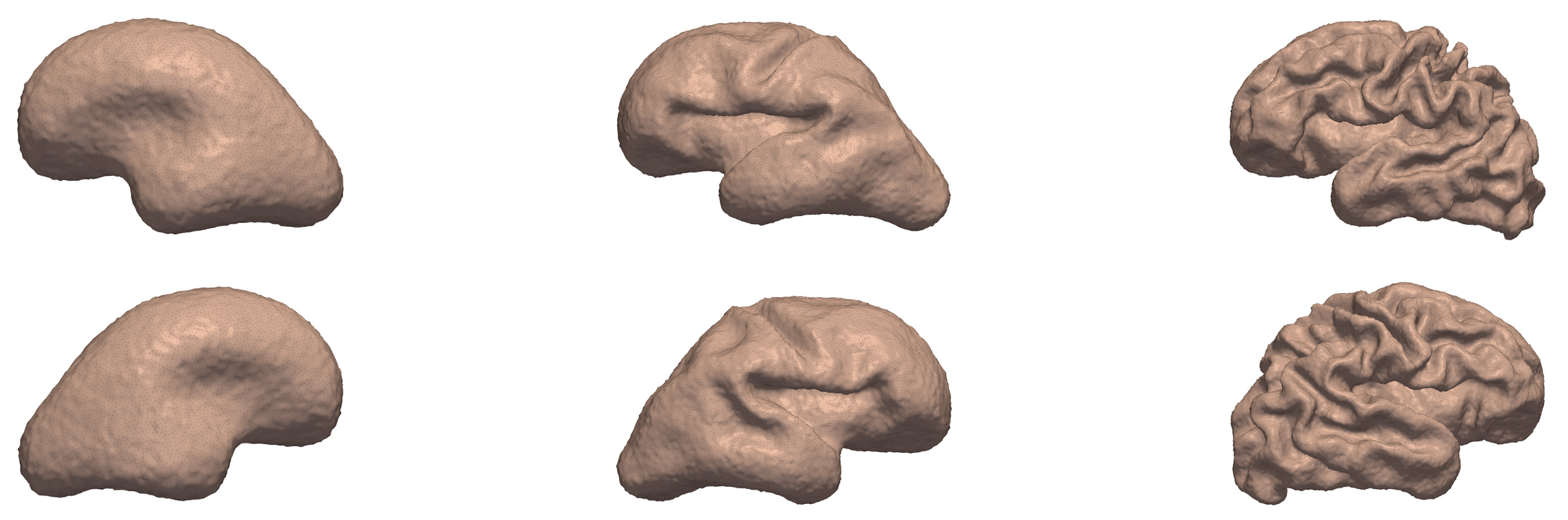}
\caption{Developmental mesh series: gestational weeks~21.86, 28.00 and~33.86 (columns); left hemisphere (top), right (bottom). Full pipeline at $h=0.6$~mm, lateral view. A cross-sectional series of three subjects, not one brain over time.}
\label{fig:dev_series}
\end{figure}

Applied to $194$ subjects spanning gestational weeks $21$--$38$ (Fig.~\ref{fig:cohort}), the pipeline holds worst-element quality below the $\qmax<0.6$ target across the entire range {($\qmax=0.46$--$0.56$)}, and the tetrahedron count grows as the cortex develops.
Quality is therefore independent of brain size and fold complexity: the same default settings yield a solver-ready mesh at every gestational age without per-subject tuning.
The developmental series in Fig.~\ref{fig:dev_series} shows the corresponding meshes to be anatomically faithful across this period, from the smooth GA-22 cortex to the deeply folded GA-34 cortex, in both hemispheres.

\subsection{Effect of resolution}
\label{sec:res-resolution}

Varying the cell size $h$ trades mesh size against resolution while leaving quality essentially unchanged (Table~\ref{tab:h_role}): the tetrahedron count scales as approximately $h^{-3}$, whereas $\bar q$ and $\qmax$ are stable across the range. The cell size therefore acts as a free density control, allowing the mesh to be refined for accuracy or coarsened for speed without compromising solver-readiness.
\begin{table}[t]
\caption{Effect of cell size $h$ on three subjects (left hemisphere, ico-6 input). Element count scales as ${\sim}h^{-3}$ while quality stays flat (lower $q$ is better).}
  \label{tab:h_role}
  \centering
  \small
  \begin{tabular}{@{}l@{\quad\quad}l@{\quad}c@{\quad}r@{\quad\quad}r@{\quad\quad}c@{\quad\quad}c@{}}
    \toprule
    Subject & GA (wk) & $h$ (mm) & Nodes & Tets & $\bar q$ & $\qmax$ \\
    \midrule
    \texttt{sub-CC01204XX08} & 21.86 & 0.4 & 194{,}632 & 1{,}101{,}843 & 0.128 & 0.550 \\
                             &       & 0.6 & 59{,}174  & 326{,}083     & 0.129 & 0.462 \\
                             &       & 0.8 & 25{,}722  & 137{,}933     & 0.130 & 0.458 \\
    \midrule
    \texttt{sub-CC00932XX17} & 28.00 & 0.4 & 458{,}388 & 2{,}612{,}132 & 0.128 & 0.522 \\
                             &       & 0.6 & 139{,}143 & 773{,}954     & 0.128 & 0.504 \\
                             &       & 0.8 & 60{,}114  & 326{,}436     & 0.129 & 0.508 \\
    \midrule
    \texttt{sub-CC01019XX13} & 33.86 & 0.4 & 955{,}059 & 5{,}419{,}671 & 0.128 & 0.540 \\
                             &       & 0.6 & 290{,}248 & 1{,}604{,}834 & 0.129 & 0.523 \\
                             &       & 0.8 & 125{,}749 & 677{,}789     & 0.129 & 0.516 \\
    \bottomrule
  \end{tabular}
\end{table}

{Worst-element quality was assessed on three representative subjects (GA-22, GA-28, GA-34; $h=0.6$~mm), giving $\qmax = 0.462$--$0.523$ under the measure \texttt{tet\_qmetric\_volume}, and, under three established literature shape measures~\cite{liujoe1994,shewchuk2002}, $\qmax = 0.339$--$0.390$ ($1-{}$mean-ratio), $0.422$--$0.497$ ($1-{}$radius-ratio), and $0.395$--$0.428$ ($1-{}$radius-edge), confirming the generated meshes are validated by multiple independent quality criteria, not \texttt{tet\_qmetric\_volume} alone.}

\subsection{A pipeline mesh sustains a folding simulation}
\label{sec:res-sim}

\begin{figure}[t]
\centering
\includegraphics[width=\textwidth]{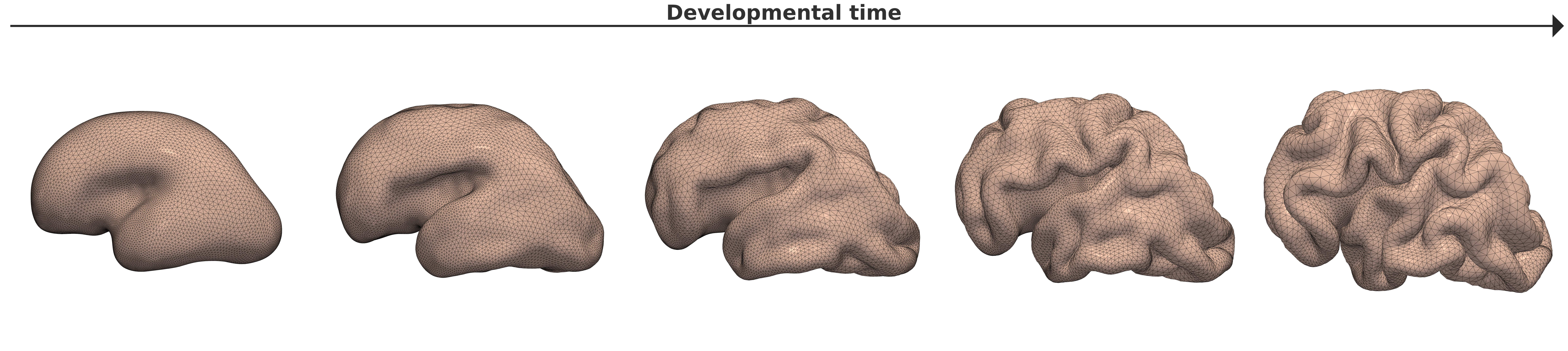}
\caption{A pipeline mesh sustains a stable folding simulation. From a smooth GA-22 cortical surface (left), the explicit morphoelastic growth model runs to a folded state (right) with no element inversion and worst-element quality below target throughout.
The folds have realistic wavelength; the resulting morphology is not claimed to match a real brain, and is a simulation result, not the imaged subject.
{The mesh was prepared with the pipeline's optional post-mesh boundary smoothing, required by the free-boundary formulation used here, followed by re-projection and meshtool re-cleaning.}}
\label{fig:sim}
\end{figure}

Finally, a mesh taken directly from the pipeline was used, without {manual repair}, to initialise an explicit morphoelastic growth simulation of a real fetal subject (Fig.~\ref{fig:sim}){, computed with our own implementation of the BrainGrowth solver of Wang et al.~\cite{wang2021}, which follows the same formulation}.
{This solver advances a growing free surface, so the residual faceting that Delaunay refinement leaves on the mesh boundary perturbs the very surface whose instability produces the folds.
The pipeline therefore provides an optional post-mesh boundary smoothing, together with the two cleaning passes it requires, and the mesh shown here was prepared with it, at the same $h=0.4$~mm resolution and smoothed input surface reported in Table~\ref{tab:h_role}. 
On this subject the step removes about $32\%$ of the boundary roughness, from $0.191$ to $0.130$, while worst-element quality improves from $\qmax=0.550$ to $0.534$ and the boundary remains within $0.173$~mm of the input surface on average ($0.434$~mm at worst), inside the $0.5$~mm isotropic resolution of the source imaging. The re-cleaning is not optional: smoothing alone leaves $\qmax=0.921$, which no explicit solver will accept. Target solvers that do not resolve a free boundary can bypass the step, as they can stages 3, 4 and~6.}
The run completed with no element inversion and worst-element quality below target throughout, producing folds of realistic wavelength from an initially smooth cortex; we make no claim that the morphology matches the individual brain. This confirms the meshes are directly usable for large-deformation folding simulation, not merely well-scored by the metric.

\section{Discussion and conclusion}
\label{sec:discussion}

We have presented the first fully automated route from an individual cortical surface to a quality-controlled tetrahedral mesh for large-deformation simulation, and shown that a mesh taken straight from it sustains a stable folding run, without manual intervention.
{Our pipeline evaluates and controls the worst-quality element rather than relying on mean mesh quality. This is essential because a single degenerate element can halt an explicit solver, even when the mesh has good average quality.}
Where a prior patient-specific fetal pipeline required manual re-meshing in a third of cases~\cite{alenya2022}, ours requires none, and its output holds $\qmax<0.6$ across the full period of cortical folding.

{A mesh that a solver will accept differs from one that merely looks correct. Five requirements govern that difference, and mesh generators enforce none of them: the input vertices are discarded and replaced during refinement, so per-vertex fields must be re-interpolated rather than indexed; a boundary-face list must be reconstructed from the volume connectivity; surface nodes must be ordered ahead of interior nodes for surface-indexed force arrays to resolve; and the per-element node ordering and signed-volume convention must match those the solver assumes. Each is invisible at mesh-generation time and appears only as a corrupted simulation, in which the run completes and looks plausible while the mechanics are wrong. CORTET enforces all five as part of a solver-ready export, which is what allows a generated mesh to be used without inspection or repair.}

\medskip\noindent\textbf{Limitations.} {Section~4.4 shows that a mesh meeting the criterion sustains a folding simulation to completion. The criterion $q_{\max}<0.6$ therefore certifies that a mesh is numerically \emph{usable}, in the sense that it does not stall the solver, and not that a simulation run on it is \emph{accurate}. Establishing the element size and quality required for a converged folding solution would need a mesh-convergence study against a reference solution.}
The target is also a strong default rather than an absolute guarantee: it was met across the developmental and resolution experiments reported here, while at the latest gestational ages deep, narrow sulci are the binding constraint on refinement, where reducing the cell size restores the target at the cost of a larger mesh. The pipeline generates a single-surface mesh, single-region, well suited to cortical folding but needing extension for multi-tissue applications. Validation draws on a single dataset and reconstruction pipeline, and pathological geometries were not specifically tested.

\medskip\noindent\textbf{Outlook.}
With the meshing bottleneck removed, mechanical parameters can be interpolated onto the volume mesh to drive spatially varying simulations, and quality-controlled meshing at cohort scale makes population studies of folding mechanics feasible, opening a route to simulation-derived mechanical biomarkers.
Scaling subject-specific folding simulations across a cohort is the natural next step toward mechanical digital twins of the developing brain, for which reliable, automated meshing is a prerequisite.
\subsubsection*{Acknowledgements.}
The authors would like to thank the following funding sources: Heart of Racing to K.R.L. and E.C.R.

\subsubsection*{Code availability.}
The CORTET pipeline is available at \url{https://github.com/Davoodshahsavari1992/-CORTET-CORtical-TETrahedral-meshing-}.
{The release covers the complete meshing pipeline as described here: surface smoothing, tetrahedral mesh generation, quality optimisation, solver-ready export, and the optional post-mesh boundary smoothing with its re-cleaning and verification steps.
The finite-element solver is not part of this release. The folding simulation in Sec.~\ref{sec:res-sim} uses our own implementation of the BrainGrowth model of Wang et al.~\cite{wang2021}, one of the few openly available solvers for cortical folding mechanics; its reference implementation is at \url{https://github.com/rousseau/BrainGrowth}.}
\bibliographystyle{spmpsci}
\bibliography{references}

\end{document}